\documentclass[aps,pra,twocolumn,showpacs,superscriptaddress]{revtex4-1}

\usepackage{amsfonts,mathrsfs,amsmath,amsthm,amssymb,bbold}                   
\usepackage{epsfig}
\usepackage{bm,bigints}
\usepackage{xfrac}
\usepackage{tabulary}
\usepackage{color}
\usepackage[dvipsnames]{xcolor}
\usepackage{graphicx}
\usepackage{hf-tikz}

\usepackage{listings}
\usepackage[utf8]{inputenc}
\usepackage{listings,xcolor}
\newcommand{\la}{\langle}
\newcommand{\ra}{\rangle}
\newcommand{\mc}{\mathcal}
\newcommand{\Omg}{\Omega}

\newcommand{\xm}{\textcolor{teal}{\text{x}\_}}
\newcommand{\xn}{\textcolor{teal}{\text{x}}}

\newcommand{\xTn}{\textcolor{teal}{\text{x2}}}

\newcommand{\sn}{\color{teal}{\sigma}}
\newcommand{\sm}{\color{teal}{\sigma\_}}

\newcommand{\bxOn}{\textcolor{blue}{\text{x1}}}
\newcommand{\bxTn}{\textcolor{blue}{\text{x2}}}
\newcommand{\bxTrn}{\textcolor{blue}{\text{x3}}}
\newcommand{\bxFn}{\textcolor{blue}{\text{x4}}}

\newcommand{\bsOn}{\color{blue}{\sigma1}}
\newcommand{\bsTn}{\color{blue}{\sigma2}}
\newcommand{\bsTrn}{\color{blue}{\sigma3}}
\newcommand{\bsFn}{\color{blue}{\sigma4}}

\newcommand{\am}{\textcolor{teal}{\text{a}\_}}
\newcommand{\an}{\textcolor{teal}{\text{a}}}
\newcommand{\emm}{\textcolor{teal}{\text{e}\_}}

\newcommand{\vm}{\textcolor{teal}{\text{vec}\_}}
\newcommand{\vn}{\textcolor{teal}{\text{vec}}}

\begin{document}  
\title {\bf Quantum backaction effects
in sequential measurements}

\author{ Le Bin Ho}
\thanks{Electronic address: 
binho@fris.tohoku.ac.jp}
\affiliation{Frontier Research Institute 
for Interdisciplinary Sciences, 
Tohoku University, Sendai 980-8578, Japan}

\affiliation{Department of Applied Physics, 
Graduate School of Engineering, 
Tohoku University, Sendai 980-8579, Japan}

\date{\today}

\begin{abstract}
Quantum backaction refers to the disturbance of a quantum system caused by measuring it. In sequential measurements, this effect can accumulate and become significant, leading to nontrivial modifications of the system state and the measurement results. This paper explores the ways in which quantum backaction can manifest in sequential measurements, including the role of measurement strength and the nature of the measurement process. The paper highlights the insight quantum foundation and the implications for quantum measurement and information processing.
\end{abstract}

%
%
\maketitle

\section{Introduction and 
preliminary definitions}
\label{seci}

Quantum mechanics predicts that
measuring a quantum state 
results in its collapse and 
reveals hidden properties
of the measuring system
\cite{wheeler_zurek_2014,david}. 
This fundamental aspect 
is known as the
``measurement problem," 
implying the state changes 
through measurements. 
The von Neumann mechanism 
explains the changing state 
via quantum backaction (QBA) 
caused by a quantum probe 
interacts with the measuring 
system~\cite{vonNeumann2018}. 
This effect is evident in 
sequential measurements, 
where later measurements 
are disturbed by the QBA 
of previous ones
\cite{Busch1990,doi:10.1063/1.1407837}. 
This causal process was widely 
explored in recent works, 
including sequential weak measurements
\cite{PhysRevA.76.062105,
PhysRevLett.117.170402,
PhysRevA.96.052123,
Brodutch2017,
PhysRevA.98.012117,
Kim2018,Chen:19,Pfender2019}, 
reduction of QBA effects
\cite{PhysRevA.51.2459,Wueaav4944,
PhysRevLett.125.210401,Shomroni2019,
Moller2017,PhysRevX.2.031016},
and manipulation of qubits \cite{Blok2014}.
Other works also include violating 
noncontextual realism \cite{doi:10.1063/1.3567466}
and extracting maximal knowledge \cite{Nagali2012} 
through sequential measurements, which are noteworthy.

In some cases, 
later measurements can noncasually 
affect the system's behaviors in the past. 
For example, in a delayed-choice 
quantum eraser experiment, 
a measurement made after 
photon detection can affect its past behavior
\cite{PhysRevLett.84.1}.
Another vital example is 
postselection measurements,
where the measurement results at present 
are affected by future postselections 
\cite{PhysRev.134.B1410,PhysRevLett.60.1351}.
Yet, studying 
the QBA caused by subsequent measurements 
is still limited, e.g.,
the quantum trajectory
in continuous measurements
\cite{PhysRevA.96.062131,
PhysRevLett.114.090403,
Ho2019,PhysRevLett.112.180402},
quantum foundation
\cite{PhysRevX.8.031013,
PhysRevLett.115.180407,
PhysRevA.88.042110},
uncertainty relations 
\cite{Bao2020,PhysRevA.104.022204},
and quantum measurements
\cite{Bao2020n,CHANTASRI20211}.
Remarkably, Tsang et al. observed 
sub-Heisenberg uncertainty via past measurements 
\cite{PhysRevA.79.053843,PhysRevA.80.033840}.
Similarly, 
Bao et al. experimentally observed the beyond 
Heisenberg uncertainty
by using past quantum states 
of preparation and postselection~\cite{Bao2020}.
These works push for scrutinizing 
the QBA in sequential measurements.

This work examines QBA effects 
in sequential measurements 
of two observables $\bm A$ and $\bm B$
in a given system,
where $\bm B$ is measured after $\bm A$.
The system is first coupled 
with a quantum probe $\mathcal{P}_1$ 
to measure $\bm A$, 
then with a probe $\mathcal{P}_2$ to measure $\bm B$. 
The probes initialize as Gaussian states  
with zero mean and are measured 
on continuous variables, e.g., $\bm x_{1,2}$.
We discuss the QBA caused by the first measurement 
affecting the second via the change in variance 
upon measurement of $\bm B$,
and the QBA caused by the second measurement 
affecting the first via the change in variance of $\bm A$.
We conduct sequential measurements 
in two models: {\it joint measurement} and 
{\it conditional measurement} (refer to Fig.~\ref{fig1}).
In the joint measurement model, 
the measuring system couples to two
quantum probes,
and the joint state is measured afterward.
We find that only the first measurement 
causally affects the second one in this model.
In the conditional measurement model, 
measurements are conditioned on each other, 
presenting QBA effects for both observables, 
including causal and noncausal disturbances.
We demonstrate these observations 
using a spin system, and uncover 
a lower bound for the stronger uncertainty 
relation proposed by Maccone and Pati 
\cite{PhysRevLett.113.260401},
enabling high-precision measurements.
We finally present a general framework 
for sequential $N$ measurements. 

QBA effects impede certain quantum 
technologies like quantum computing 
and metrology. Understanding and 
controlling QBA in sequential 
measurements could improve the quantum 
measurement foundation, increase 
precision, and offer noise mitigation techniques.

The paper is structured as follows. 
Section~\ref{secii} presents 
the sequential measurements framework, 
including the joint and conditional measurement models. 
Illustrative results are shown in Section~\ref{seciii}, 
and the general framework for 
$N$-sequential measurements 
is provided in Section~\ref{seciv}. 
The paper is summarized in Section~\ref{secv}.

\section{Sequential measurements}\label{secii}

\begin{figure} [t]
\centering
\includegraphics[width=8.6cm]{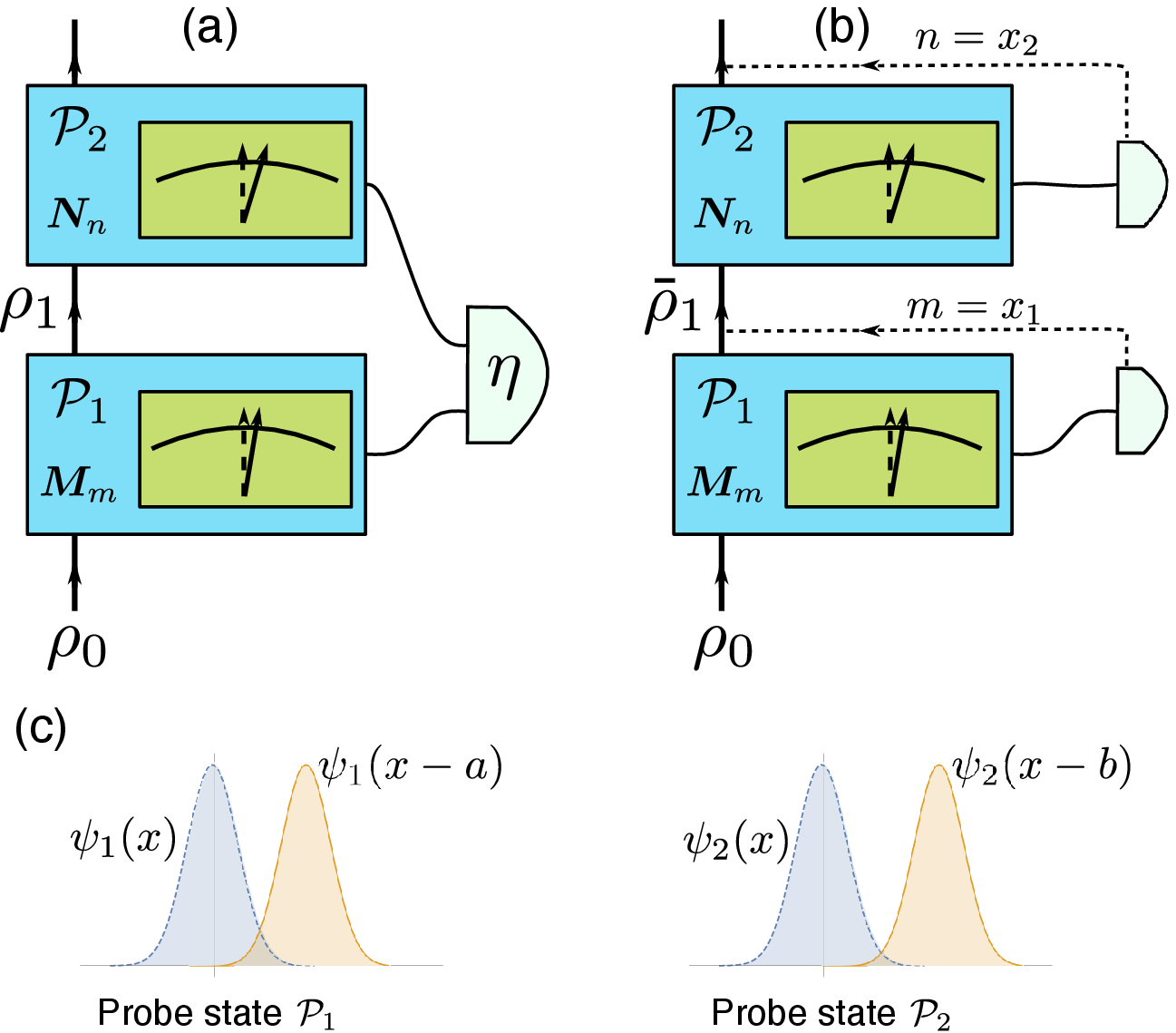}
\caption{
Schematic of sequential measurements:
(a) joint measurement model,
(b) conditional measurement model.
System $\mc{S}$ is initially prepared 
in $\rho_0$.
The measurement of $\bm A$ 
in the system is
done via a set of 
measurement operators
$\bm M_m$
in the first 
quantum probe
 $\mc{P}_1$.
The initial state of the probe
is $\psi_1(x)$,
and it transforms to $\psi_1(x-a)$ 
after the interaction (figure c),
which imprints the information about the 
eigenvalue $a$ of the operator $\bm A$. 
As a QBA effect, the interaction 
disturbs system $\mc{S}$
and transforms the system state 
to $\rho_1$ in the 
joint measurement model (a)
or conditional $\bar \rho_1$ 
in the conditional measurement model (b).
Likewise, the measurement of  
$\bm B$ is taken in the same manner. 
It is given by a set of 
measurement operators $\bm N_n$,
where the initial state of 
the second probe
$\psi_2(x)$ transforms to $\psi_2(x-b)$,
which corresponds to the eigenvalue $b$ 
of operator $\bm B$.
In the conditional measurement model,
we only keep the results when
the outcomes $m = x_1$
and $n = x_2$, while the others are discarded.
}
\label{fig1}
\end{figure} 

For a given system $\mc{S}$,
we consider a measurement of 
$\bm A$ followed by
an incompatible measurement of $\bm B$,
where $\bm A$ and $\bm B$
are selfadjoint operators.
These measurements
are performed by coupling $\mc{S}$
with two quantum probes
$\mc{P}_1$ 
and $\mc{P}_2$, respectively.
The measurement scheme 
is schematically shown in 
Fig.~\ref{fig1}.
Concretely, let us consider 
an atom-light interaction scheme 
\cite{RevModPhys.82.1041},
 where $\mc{S}$ is an atomic ensemble 
that interacts with light (quantum probes). 
The interactions are governed by
\begin{align}\label{UAB}
\bm U_1 = e^{-i\bm A\otimes \bm p_1}, \text{ and }\, 
\bm U_2 = e^{-i\bm B\otimes \bm p_2},
\end{align}
where $\bm p_1$ and $\bm p_2$ are 
canonical momentum operators 
of $\mc{P}_1$ and $\mc{P}_2$,
respectively.
The initial state for each probe is
described by a Gaussian 
wave function with a mean of zero, such as
$|\psi_i(x)\ra = 
\int
\psi_i(x) |x\ra \ {\rm d}x$, 
where 
\begin{align}\label{GaussianAB}
\psi_i(x) = 
\Bigl(\dfrac{1}{2\pi\sigma_i^2}\Bigr)^{1/4}
\exp\Bigl(-\frac{x^2}{4\sigma_i^2}\Bigr),
\end{align}
is the quadrature distribution 
of the input probe state, 
$i = 1, 2$ correspond to
$\mc{P}_1$ and $\mc{P}_2$, 
respectively,
$|x\ra$ is a continuous eigenstate of
the
canonical position operator $\bm x$, 
and $\sigma_i$ is the width of 
the wave function,
which stands for the measurement strength:
$\sigma_i \to 0$ is a strong measurement, while
$\sigma_i \to \infty$ is a weak measurement.
Such interaction models
were wildely implemented in 
atom-light interaction concepts
\cite{RevModPhys.82.1041,
Julsgaard_2003,
Colangelo2017, 
PhysRevA.96.063402,
Atature2007,Vasilyev_2012,
doi:10.1063/1.2721380,
Bao2020,
PhysRevLett.124.110503,
PhysRevA.70.052324,
PhysRevLett.102.125301,
PhysRevA.105.052228},
and extensively applied for  
simultaneous measurement of 
spin angle and amplitude 
\cite{Colangelo2017},
beyond Heisenberg
uncertainty \cite{Bao2020},
and error-disturbance uncertainty 
in Faraday measurement
\cite{PhysRevA.105.052228}.

Below, we examine the QBA effects
by analyzing the variances of measured operators 
in a joint measurement model
and a conditional measurement model.

\subsection{Joint measurement model}
We first consider the joint measurement approach
shown in Fig.~\ref{fig1} (a).  
Assume the initial joint state  of 
$\mc{S}\otimes\mc{P}_1\otimes\mc{P}_2$ is 
\begin{align}\label{eq:rhoint}
\rho_{\rm in} = \rho_0 \otimes 
|\psi_1(x)\ra\la\psi_1(x)|
\otimes |\psi_2(x)\ra\la\psi_2(x)|,
\end{align}
where $\rho_0$ is the initial system state.
The joint state transforms to 
$\rho_{\rm fin} = \bm U_2\bm U_1\rho_{\rm in}
\bm U_1^\dagger \bm U_2^\dagger$.
Inserting identity operators 
$\bm I_1 = \int |m\ra\la m| {\rm d}m$
and $\bm I_2 = \int |n\ra\la n| {\rm d}n$
into the probes, 
we obtain 
\begin{align}\label{fin_state}
\rho_{\rm fin} = 
\int\bm N_{n}
\bm M_{m}\rho_0\bm M^\dagger_{m'}
\bm N_{n'}^\dagger
\otimes |m\ra\la m'| 
\otimes |n\ra\la n'|
\ {\rm d}^4\bm{t},
\end{align}
where $\bm t$ represents 
a 4-tuple $(m, m', n, n')$
and ${\rm d}^4\bm t$ is 
a 4-dimensional volume differential
(see App.~\ref{appA}).
The measurement operators $\bm M$ and $\bm N$ 
for the first and second measurements are given by 
\begin{align}\label{MN}
\notag \bm M_{m}
&=\int
\psi_1(m-a)|a\ra\la a|{\rm d}a, \\
\bm N_{n}
&=\int
\psi_2(n-b)|b\ra\la b|{\rm d}b,
\end{align}
where $|a\ra$ is an eigenstate of 
$\bm A$ per an eigenvalue $a$, 
and similar for $\bm B$.
These measurement operators 
satisfy the completeness equations
$\int \bm M_{m}^\dagger \bm M_{m} {\rm d}m
= \bm I$
and $\int \bm N_{n}^\dagger \bm N_{n} {\rm d}n
= \bm I$.
The final
state of 
the two probes 
is given by
$\eta  = 
{\rm Tr}_{\mc{S}} [\rho_{\rm fin}] 
$, where Tr$_\mc{S}[*]$ 
is a partial trace w.r.t. 
system $\mc{S}$.
It explicitly yields (see App.~\ref{appB})
\begin{widetext}
\begin{align}\label{joint_state2} 
\eta  = 
\iiint \la b|a\ra
\la a|\rho_0| a'\ra
\la a'|b\ra\times 
|\psi_1(x-a)\ra
\la\psi_1(x-a')|\otimes |\psi_2(x-b)\ra
\la\psi_2(x-b)|
\ {\rm d}a {\rm d}a' {\rm d}b.
\end{align}
\end{widetext}

After the interaction, 
we measure the position
operator $\bm x_1$ 
in pointer $\mc{P}_1$
and get 
the expectation value as
\begin{align}\label{fe}
\la\bm x_1\otimes \bm I_2\ra
= {\rm Tr}_{\mc{P}_1\mc{P}_2}
\bigl[(\bm x_1\otimes \bm I_2)\eta\bigr]
={\rm Tr}_{\mc{S}}
[\bm A\rho_0].
\end{align}
The variance of observed values 
upon measurement of $\bm x_1$
is given by  
\begin{align}\label{var1}
\notag {\rm Var}(\bm x_1)_{\eta} & \equiv
\la(\bm x_1\otimes \bm I_2)^2\ra
- \la\bm x_1\otimes \bm I_2\ra^2\\
& = \sigma_1^2+{\rm Var}(\bm A)_{\rho_0},
\end{align}
which is the spread of the observed values 
upon measurement of $\bm x_1$
concerning $\eta$
(shorthanded as the variance of $\bm x_1$).
Here, ${\rm Var}(\bm A)_{\rho_0}
= {\rm Tr}_{\mc{S}}[\bm A^2\rho_0] - 
({\rm Tr}_{\mc{S}}[\bm A\rho_0])^2$
is the variance of 
$\bm A$
concerning 
$\rho_0$
(see App.~\ref{appC1}).
Experimentally, one can infer 
${\rm Var}(\bm A)_{\rho_0} = 
{\rm Var}(\bm x_1)_\eta-\sigma_1^2$,
where $\sigma_1^2$
stands for shot noise fluctuations
\cite{Bao2020}.

The QBA effect caused by  the first measurement 
can be informed 
by the change in variance 
of the second measurement.
To evaluate this 
QBA, we examine the 
variance of 
$\bm x_2$
in the second measurement
as follows
\begin{align}\label{varB}
\notag{\rm Var}(\bm x_2)_\eta &\equiv 
\bigl\la(\bm I_1\otimes \bm x_2)^2\bigr\ra
-\bigl\la\bm I_1\otimes \bm x_2\bigr\ra^2,\\
&=\sigma_2^2 + {\rm Var}(\bm B)_{\rho_1},
\end{align}
where the expectation value is
$\la\bm I_1\otimes \bm x_2\ra
= {\rm Tr}_{\mc{P}_1\mc{P}_2}
[(\bm I_1\otimes \bm x_2)\eta]$,
and
\begin{align}\label{varB'}
{\rm Var}(\bm B)_{\rho_1} =
{\rm Tr}_{\mc{S}}[\bm B^2\rho_1]
-({\rm Tr}_{\mc{S}}[\bm B\rho_1])^2,
\end{align}
is the variance of $\bm B$
concerning $\rho_1$. 
Here, $\rho_1
= {\rm Tr}_{\mc{P}_1}
[\bm U_1(\rho_0 \otimes 
|\psi_1(x)\ra\la\psi_1(x)|)
\bm U_1^\dagger]$ is 
the system state after
interacting with $\mc{P}_1$.
Straightforwardly, it yields 
(see App.~\ref{appC2})
\begin{align}\label{rho1}
\rho_1 = \iint |a\ra\la a|
\rho_0
|a'\ra\la a'|
e^{-\frac{(a-a')^2}{8\sigma_1^2}}
{\rm d}a {\rm d}a'.
\end{align}
Obviously, if the initial state
$\rho_0$ is an eigenstate of $\bm A$,
i.e., $\rho_0 = |a_i\ra\la a_i|$,
then $\rho_1 = \rho_0$,
meaning that system $\mc{S}$
is not disturbed after 
the first measurement
\cite{PhysRevLett.125.210401}. 
In general, however, 
the system
is disturbed through 
the first measurement,
and thus, $\rho_1 \ne \rho_0$.
Hence, Eq.~\eqref{varB'} 
explicitly gives

\begin{align}\label{DelBp}
{\rm Var}(\bm B)_{\rho_1} = 
\int b^2 \la b|\rho_1|b\ra  {\rm d}b
- \Bigg(
\int  b \la b|\rho_1|b\ra  {\rm d}b
\Bigg)^2,
\end{align}
which significantly
depends on 
the first measurement
(via $\sigma_1$) for all 
$a \neq a'$.

In general, 
${\rm Var}(\bm B)_{\rho_1} 
\ne {\rm Var}(\bm B)_{\rho_0}$
due to the QBA caused by the first measurement,
where ${\rm Var}(\bm B)_{\rho_0} = 
{\rm Tr}_{\mc{S}}[\bm B^2\rho_0] - 
({\rm Tr}_{\mc{S}}[\bm B\rho_0])^2$
is the variance of 
$\bm B$ 
concerning
 $\rho_0$. 
The dependence of ${\rm Var}(\bm B)_{\rho_1}$ 
on the first measurement illustrates
an {\it in situ causal effect}, 
where later measurement results 
are affected by previous measurements.

When the 
first measurement is weak,
i.e., $\sigma_1\to \infty$, 
Eq.~\eqref{varB}
 simplifies to
${\rm Var}(\bm x_2)_\eta  
= \sigma_2^2+{\rm Var}(\bm B)_{\rho_0}$,
which is not affected by the first measurement.
This means
the QBA is eliminated 
\cite{PhysRevA.97.012122,
Abbott2019anomalousweakvalues}.
However, the cost is that 
the fluctuation of the first 
measurement will increase,
i.e., ${\rm Var}(\bm x_1)_\eta$ is large~
\cite{PhysRevA.97.012122,
Aharonov2009}. 
Discussing this increasing fluctuation
is out of the scope,
where we instead focus on
the mutual influence in
sequential measurements.

\subsection{Conditional measurement model}

In this approach, assume that 
we first measure $\bm x_1$ in
$\mc{P}_1$ and get an outcome 
$x_1$.
After the first measurement,
the initial system state $\rho_0$
transforms into
an unnormalized state
\begin{align}\label{sysstate}
\notag\bar{\rho}_1(x_1) & = 
\bm M_{x_1}\rho_0\bm M^\dagger_{x_1}\\
&=\iint |a\ra\la a|\rho_0|
a'\ra\la a'|\psi_1(x_1-a)\psi_1^*(x_1-a') 
{\rm d}a{\rm d}a'.
\end{align}
Here, we used $\bar{\rho}_1(x_1)$
to specify that the system state 
is given according to 
the outcome $x_1$ of the first measurement
(see, for example,
Chap. II in Ref.~\cite{nielsen_chuang_2010}).
After the measurement 
of $\bm x_2$ with an outcome
$x_2$, the system state becomes
$\bar{\rho}_2(x_1, x_2) \propto 
\bm N_{x_2}\bar\rho_1(x_1)\bm N^\dagger_{x_2}$.
For convenience, we introduce an effective matrix 
$\bm E(x_2) = \bm N^\dagger_{x_2}
\bm N_{x_2}$, which is the past quantum state
\cite{PhysRevLett.111.160401}.

We examine the QBA effect 
caused by the first measurement, 
i.e., the outcome of the second
measurement 
is affected by the first measurement.
We consider the conditional probability for 
obtaining the outcome $\bm x_2 = x_2$ 
conditioned on $\bm x_1 = x_1$ as 
(see App.~\ref{appD})
\begin{align}\label{cp}
\notag p(x_2|x_1) &= 
\dfrac{
{\rm Tr}\bigl[\bar{\rho}_1(x_1)
\bm E(x_2)\bigr]}
{\bigintssss
{\rm Tr}\bigl[\bar{\rho}_1(x_1)
\bm E(x'_2)\bigr] {\rm d}x'_2}\\
&=\dfrac{
\bigintssss \big|\psi_2(x_2-b)\big|^2\la b|\bar{\rho}_1(x_1)
| b\ra\ {\rm d}b}
{\bigintssss\Bigl[\bigintssss
\big|\psi_2(x'_2-b)|^2\la b|\bar{\rho}_1(x_1)
|b\ra\ {\rm d}b\Bigr] {\rm d}x'_2},
\end{align}
where $p(x_2|x_1)$ is
the shorthand of 
$p(\bm x_2 \!\!= \!x_2|\bm x_1 \!\!=\! x_1)$.
Then, the expectation value 
of $\bm x_2$ conditioned on the 
outcome $\bm x_1 = x_1$
reads
\begin{align}\label{expx2}
E(\bm x_2|\bm x_1\!\!=\! x_1) = 
\int x_2\
p(x_2|x_1)
{\rm d}x_2,
\end{align}
where `$E$' stands 
for $\la\cdots\ra$ as an expectation value.
Generally, this expectation value 
is affected by the QBA 
caused by the first measurement.
The conditional variance 
of $\bm x_2$ gives 
\begin{align}\label{condx2}
\notag {\rm Var}(\bm {x}_2|\bm x_1 \!\!= \!x_1)&= 
E(\bm x_2^2|\bm x_1 \!\!=\!x_1) -
[E(\bm x_2|\bm x_1 \!\!=\! x_1)]^2
\\
&= \sigma_2^2 + {\rm Var} (\bm B|\bm A),
\end{align}
where ${\rm Var} (\bm B|\bm A)$
is the variance of 
$\bm B$ 
conditioned on the presence of $\bm A$. 
In indirect measurements, 
this variance is informed by 
Eq.~\eqref{condx2} as
\begin{align}\label{eq:var21} 
{\rm Var} (\bm B|\bm A)
= {\rm Var}(\bm {x}_2|\bm x_1 
\!\!=\!x_1)-\sigma_2^2.
\end{align}
Generally, it depends on 
the strength  
of the first and second measurements.
The QBA effect appears when 
${\rm Var}(\bm B|\bm A)
\ne {\rm Var}(\bm B)_{\rho_0}$.

Likewise, 
the second measurement 
can noncausally affect back
the first measurement.
To evaluate the noncausal backaction,
we consider the conditional probability
for obtaining $\bm x_1 = x_1$
conditioned on $\bm x_2 = x_2$
\begin{align}\label{cp2}
\notag p(x_1|x_2) &= 
\dfrac{
{\rm Tr}\bigl[\bar{\rho}_1(x_1)
\bm E(x_2)\bigr]}
{\bigintssss
{\rm Tr}\bigl[\bar{\rho}_1(x'_1)
\bm E(x_2)\bigr] {\rm d}x'_1},\\
&=\dfrac{
\bigintssss \big|\psi_2(x_2-b)\big|^2\la b|\bar{\rho}_1(x_1)
| b\ra\ {\rm d}b}
{\bigintssss\Bigl[\bigintssss
\big|\psi_2(x_2-b)\big|^2\la b|\bar{\rho}_1(x'_1)
|b\ra\ {\rm d}b\Bigr] {\rm d}x'_1}.
\end{align}
The conditional expectation value 
and conditional variance
are given by 
$E(\bm{x}_1|\bm x_2 \!\!=\!{x_2})
= \int x_1 p(x_1|x_2) {\rm d}x_1
$ 
and ${\rm Var}(\bm {x}_1|\bm x_2\!\!=\! x_2) = 
E(\bm{x}_1^2|\bm x_2 \!\!=\!{x_2})
-\bigl[E(\bm{x}_1|\bm x_2 \!\!=\!{x_2})]^2$,
respectively. 
Similarly, the variance 
of $\bm A$ conditioned on
$\bm B$ 
can be informed from
\begin{align}\label{caonAb}
{\rm Var}(\bm A|\bm B)
= {\rm Var}(\bm x_1|\bm x_2 \!\!=\!x_2) - \sigma_1^2.
\end{align}
Here, 
${\rm Var}(\bm A|\bm B)
\ne {\rm Var}(\bm A)_{\rho_0}$
is the result of the noncausal QBA effect
caused by the second measurement.
Equations~(\ref{eq:var21}, \ref{caonAb}) explicitly 
show the impacts of 
the first measurement 
on the second measurement 
and vice versa, respectively.

Intuitively, 
the conditional state $\bar{\rho}_1$
interferes with
the effective matrix  $\bm E$
in the same way 
as the interference of pre- 
and post-selection states,
which was experimentally observed 
in the resonance fluorescence
\cite{PhysRevLett.112.180402}.
The interference here 
relies on the conditional outcomes
$x_1, x_2$
(like postselection), 
allowing measurements of $\bm A$
and $\bm B$ to affect each other.
This is a purely quantum feature 
of the reciprocal past-future information
\cite{CHANTASRI20211}.
The interference can be observed 
through continuous monitoring 
of the system under different observables.
\cite{
PhysRevA.96.062131,
PhysRevLett.112.180402,
Ho2019}.

\subsection{Conditional stronger uncertainty relations} 
It is unfeasible to get precise results 
in incompatible measurements, 
whether they are simultaneous or sequential
\cite{PhysRev.34.163, 
PhysRevLett.60.2447,
PhysRevA.97.012122}.
The Heisenberg-Robertson uncertainty
relation
for two incompatible observables $\bm A$ 
and $\bm B$ can be given 
in terms of the commutator 
${\rm Var}(\bm A)
{\rm Var}(\bm B)
\ge \bigl| \frac{1}{2}\la
[\bm A, \bm B]\ra\bigr|^2$,
which is the very first investment 
on this issue \cite{PhysRev.34.163, 
PhysRevLett.60.2447}.
However, the lower bound in this relation is trivial, 
such as
it can reach zero
when the system state is an eigenstate
of either $\bm A$ or $\bm B$.
Maccone and Pati 
later proposed a stronger (nontrivial) 
uncertainty relation in terms 
of the summation of variances 
\cite{PhysRevLett.113.260401}
\begin{align}\label{eq:MUP}
{\rm Var}(\bm A)+{\rm Var}(\bm B)
\ge \max (\mc{R}_{\rm a}, \mc{R}_{\rm b}),
\end{align}
which we name as Maccone-Pati uncertainty 
relation (MPUR),
where 
\begin{align} \label{eq:MPAB}
\notag  \mc{R}_{\rm a} &= 
        \pm i\la[\bm A,\bm B]\ra
        +\bigl|\la\psi|\bm A\pm i\bm B
        |\psi^\perp\ra\bigr|^2, \\
        \mc{R}_{\rm b} &=
       \dfrac{1}{2}\bigl|\la
        \psi^\perp_{\bm A + \bm B}
        |\bm A+\bm B|\psi\ra\bigr|^2.
\end{align}
Here, $|\psi^\perp\ra$ 
is an arbitrary orthogonal state
and $|\psi^\perp_{\bm A+\bm B}
\ra \propto (\bm A + \bm B -
\la\bm A+\bm B\ra)|\psi\ra$.
The sign should be chosen 
so that $\pm i\la[\bm A,\bm B]\ra$
is positive.
The MPUR is nontrivial,
i.e., nonzero lower bound.
Such an MPUR has been experimentally
verified in Ref.~\cite{PhysRevA.93.052108}.

In this framework, 
we examine the MPUR
in terms of conditional variances,
where the summation  
${\rm Var}(\bm A)+{\rm Var}(\bm B)$ 
is replaced by ${\rm Var}(\bm A|\bm B)
+{\rm Var}(\bm B|\bm A)$.
In the following section,
we observe that 
$
{\rm Var}(\bm A|\bm B)+
{\rm Var}(\bm B|\bm A)
 < \max (\mc{R}_{\rm a}, 
 \mc{R}_{\rm b})$.
This observation is a result of the 
conditional measurement, 
which analogizes to the  
postselection measurement 
\cite{Abbott2019anomalousweakvalues,Bao2020}. 
We emphasize that 
this result does not contradict 
Eq.~\ref{eq:MUP} since it only 
applies to conditional cases.
Significantly, it allows for
getting higher measurements precision
and minimizing the trade-off between
${\rm Var}(\bm A|\bm B)$ and
${\rm Var}(\bm B|\bm A)$. 
The result is thus significant for minimizing
QBA effects in quantum measurements.

\begin{figure} [t]
\centering
\includegraphics[width=8.6cm]{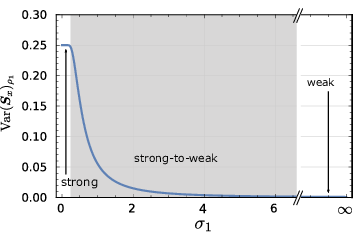}
\caption{
The variance of $\bm S_x$
after the first measurement, ${\rm Var}(\bm S_x)_{\rho_1}$.
The measurement of $\bm S_x$ is affected 
by the first measurement via $\sigma_1$.
When the first measurement is strong, 
i.e., $\sigma_1\to 0$, the variance 
${\rm Var}(\bm S_x)_{\rho_1}$ is large, 
indicating a large disturbance. 
When the first measurement is weak, i.e.,
large $\sigma_1$, ${\rm Var}(\bm S_x)_{\rho_1}$  
reduces to zero and equals to ${\rm Var}(\bm S_x)_{\rho_0}$,
which means that the system 
is not disturbed. 
In the strong-to-weak transition regime,
${\rm Var}(\bm S_x)_{\rho_1}$ reduces from 
1/4 to 0, resulting 
in a decreased disturbance.
}
\label{fig2}
\end{figure} 

\begin{figure*} [t]
\centering
\includegraphics[width=\textwidth]{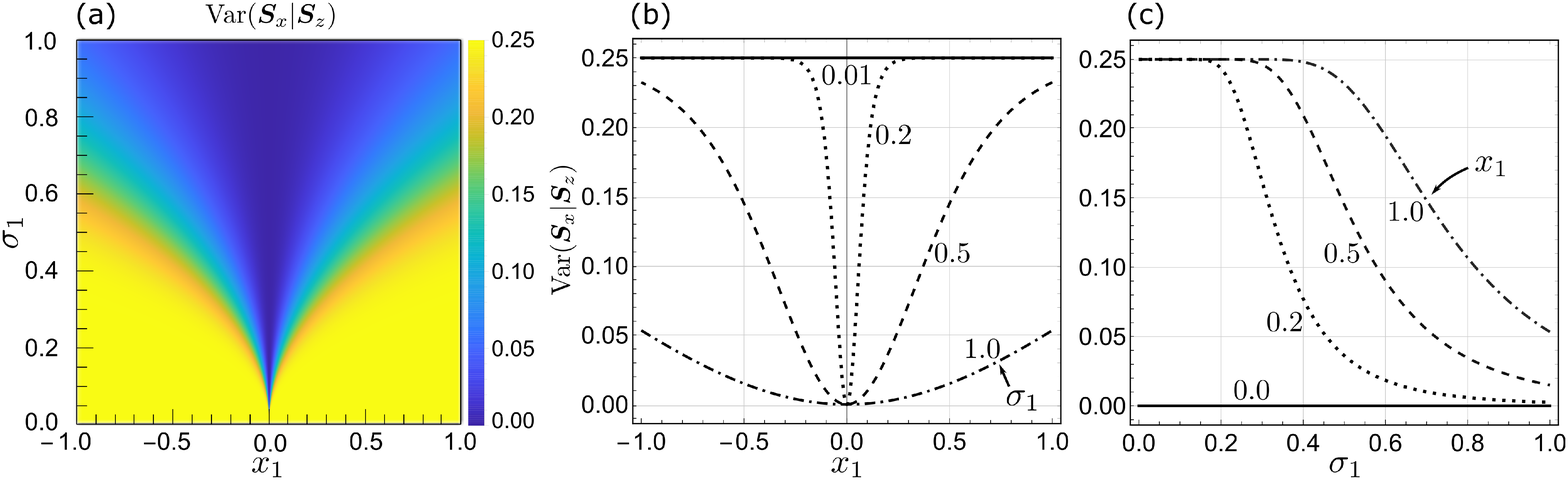}
\caption{
(a) The conditional  
variance ${\rm Var}(\bm S_x|\bm S_z)$
in Eq.~\eqref{dBA}
as a function of $x_1$ and $\sigma_1$.
When ${\rm Var}(\bm S_x|\bm S_z) = 
{\rm Var}(\bm S_x)_{\rho_0} = 0$,
there is no disturbance for the system,
such as when $x_1 = 0$.
For small $\sigma_1$ and $x_1\ne 0$, 
${\rm Var}(\bm S_x|\bm S_z)$ reaches  
1/4, the maximum disturbance 
resulting from the QBA.
For large $\sigma_1$, 
${\rm Var}(\bm S_x|\bm S_z)$
reduces to zero due to
the weak measurement.
(b) Plot of ${\rm Var}(\bm S_x|\bm S_z)$
as a function of $x_1$ for 
$\sigma_1\in\{0.01, 0.2, 0.5, 1.0\}$.
(c) Plot of ${\rm Var}(\bm S_x|\bm S_z)$
as a function of $\sigma_1$ for 
$x_1\in\{0.0, 0.2, 0.5, 1.0\}$.
}
\label{fig3}
\end{figure*} 

\section{Illustration in a spin system}\label{seciii}
We consider the sequential measurements of
a spin system 
using an atom-light interaction scheme~
\cite{RevModPhys.82.1041,
Julsgaard_2003,
Colangelo2017, 
PhysRevA.96.063402,
Atature2007,Vasilyev_2012,
doi:10.1063/1.2721380,
Bao2020,
PhysRevLett.124.110503,
PhysRevA.70.052324,
PhysRevLett.102.125301,
PhysRevA.105.052228}.
To simplify, we consider the system 
to be a single spin-1/2 and choose
$\bm A = \bm S_z = \bm \sigma_z/2$
and $\bm B = \bm S_x = \bm \sigma_x/2$,
where $\bm\sigma_i, (i = x, y, z)$
is a Pauli matrix.
The initial system state is
$\rho_0 = |+\ra\la+|$, where 
$|\pm\ra = (|\uparrow\ra \pm |\downarrow\ra)/\sqrt{2}$,
with $|\uparrow\ra$ and $|\downarrow\ra$
are the eigenstates of $\bm S_z$.
This choice gives 
${\rm Var}(\bm S_z)_{\rho_0} = 1/4$ 
and ${\rm Var}(\bm S_x)_{\rho_0} = 0$.
Additionally, we expand operators $\bm S_z$
and $\bm S_x$ into their eigenvalues and
eigenstates as follows
\begin{align}\label{eq:szsx}
\bm S_z &\equiv 
\sum_i a_i |a_i\ra\la a_i|
= \frac{1}{2}
|\uparrow\ra\la\uparrow| 
- \frac{1}{2}|\downarrow\ra
\la\downarrow|),\\
\bm S_x &\equiv \sum_i b_i |
b_i\ra\la b_i| =
\frac{1}{2}
|+\ra\la+| 
-\frac{1}{2} |-\ra\la-|).
\end{align}
Since these spectrums are discrete, 
we replace the integrals (w.r.t d$a$ and d$b$)
with the summation over the discrete eigenstates.

\subsection{Joint measurement model}
First, we calculate $\rho_1$
in Eq.~\eqref{rho1}:
\begin{align}\label{rho1i}
\notag \rho_1 &= 
\sum_{i,j \in \{\uparrow, \downarrow\}}
|a_i\ra\la a_i
|+\ra\la +|
a_j\ra\la a_j|
e^{-\frac{(a_i-a_j)^2}{8\sigma_1^2}}\\
&=\dfrac{1}{2}
\begin{pmatrix}
   1 & e^{-\frac{1}{8\sigma_1^2}} \\
   e^{-\frac{1}{8\sigma_1^2}} & 1
\end{pmatrix}.   
\end{align}
Substituting Eq.~\eqref{rho1i} 
into Eq.~\eqref{DelBp},
we obtain
\begin{align}\label{DelBpi}
\notag {\rm Var}(\bm S_x)_{\rho_1} &= 
\sum_{i\in\{+,-\}} b_i^2 \la b_i|\rho_1|b_i\ra 
- \Bigl[
\sum_{i\in\{+,-\}}  b_i \la b_i|\rho_1|b_i\ra
\Bigr]^2\\
&= \dfrac{1}{4}
\Bigl(1 -  e^{-\frac{1}{4\sigma_1^2}}\Bigr).
\end{align}
  
In Fig.~\ref{fig2}, we plot 
${\rm Var}(\bm S_x)_{\rho_1}$ 
versus the strength of the first measurement,
$\sigma_1$.
For strong measurement ($\sigma_1\ll 1$), 
${\rm Var}(\bm S_x)_{\rho_1}$ reaches
the maximum of 1/4,
indicating that the system 
is maximum disturbed by
the first measurement.
For weak measurement
($\sigma_1\to \infty$),
${\rm Var}(\bm S_x)_{\rho_1}$
decreases and approaches 0,
which means that 
the first measurement 
has minimal impact on 
the second one.
During the strong-to-weak transition regime
\cite{Piacentini2018,Pan2020}, 
${\rm Var}(\bm S_x)_{\rho_1}$ reduces from 
1/4 to 0, resulting 
in a decrease in QBA from strong 
to weak interactions.

\subsection{Conditional measurement model}
First, let us calculate $\bar\rho_1(x_1)$
in Eq.~\eqref{sysstate}:
\begin{align}\label{sysstatei}
\notag\bar{\rho}_1(x_1)
&=\sum_{i,j \in \{\uparrow, \downarrow\}} 
|a_i\ra\la a_i|\rho_0|
a_j\ra\la a_j|\psi_1(x_1-a_i)\psi_1^*(x_1-a_j) \\
&=\dfrac{1}{2}\Bigl(
\frac{1}{2\pi\sigma_1^2}\Bigr)^{1/2}
e^{-\frac{x_1^2+1/4}{2\sigma_1^2}}
\begin{pmatrix}
e^{\frac{x_1}{2\sigma_1^2}} & 1\\
1 & e^{-\frac{x_1}{2\sigma_1^2}}
\end{pmatrix}.
\end{align}
This state significantly depends on the 
outcome $x_1$ and the interaction strength
$\sigma_1$
of the first measurement.  
The conditional 
expectation values
in Eq.~\eqref{expx2} yield 
\begin{align}\label{expx2i}
E(\bm x_2|\bm x_1 \!\!=\!x_1) &= 
\dfrac{{\rm exp}({\frac{x_1}{2\sigma_1^2}})}
{1+{\rm exp}({\frac{x_1}{\sigma_1^2}})}, \\
E(\bm x_2^2|\bm x_1 \!\!=\!x_1) &= 
\dfrac{1}{4} + \sigma_2^2,
\end{align}
and the conditional variance 
of $\bm x_2$ is given by
\begin{align}\label{condx2i}
{\rm Var}(\bm {x}_2|\bm x_1\!\!=\!x_1) = 
\dfrac{1}{4} \tanh\Bigl(\dfrac{x_1}
{2 \sigma_1^2}\Bigr)^2 + \sigma_2^2.
\end{align}
Inversely, 
the variance of
$\bm S_x$ conditioned on $\bm S_z$ gives
\begin{align}\label{dBA}
{\rm Var}(\bm S_x|\bm S_z) = 
\dfrac{1}{4} \tanh\Bigl(\dfrac{x_1}
{2 \sigma_1^2}\Bigr)^2.
\end{align}
We show the result in Fig.~\ref{fig3}.
When the first measurement is strong
($\sigma_1 \to 0$), 
the conditional variance   
reaches the maximum of 1/4 
for $x_1 \ne 0$, 
indicating that the system is 
maximally disturbed by 
the first measurement
due to the QBA. 
In other words, before the first measurement,
the variance ${\rm Var}(\bm S_x)_{\rho_0} = 0$;
and after the measurement, it
reaches the maximum 
${\rm Var}(\bm S_x|\bm S_z) = 1/4$.
When the first measurement is weak
($\sigma_1\to\infty$), 
it disturbs the system less, and thus
${\rm Var}(\bm S_x|\bm S_z)$ reduces
to ${\rm Var}(\bm S_x)_{\rho_0}$.
This observation is consistent with
the joint measurement approach above,
where the first measurement 
affects the results of the second.

\begin{figure*} [t]
\centering
\includegraphics[width=\textwidth]{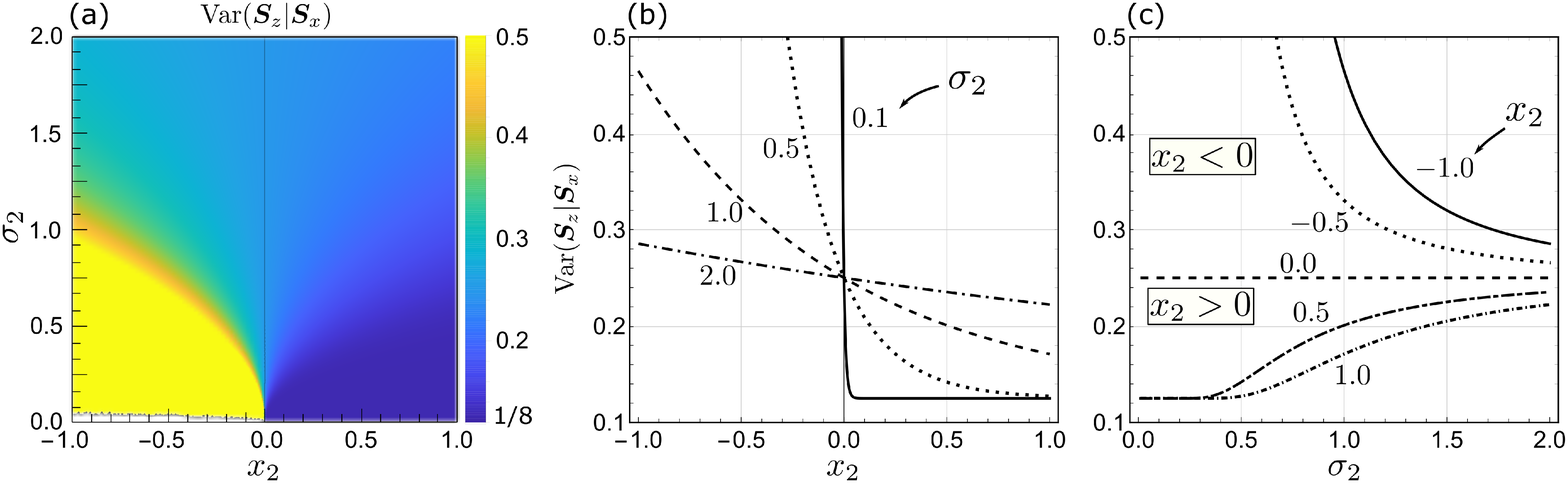}
\caption{
(a) The conditional  
variance ${\rm Var}(\bm S_z|\bm S_x)$
in Eq.~\eqref{dAB}
as a function of $x_2$ and $\sigma_2$
when $\sigma_1\to\infty$.
When ${\rm Var}(\bm S_z|\bm S_x) = 
{\rm Var}(\bm S_z)_{\rho_0} = 1/4$,
there is no disturbance for the system,
such as when $x_2 = 0$.
For small $\sigma_2$ and $x_2 < 0$, 
then ${\rm Var}(\bm S_z|\bm S_x) > 1/4$, 
which implies the QBA caused by 
the second measurement affecting
the first. Here, 
${\rm Var}(\bm S_z|\bm S_x) \to \infty$
when $\sigma_2\to 0$,
however, we show the result up to 
${\rm Var}(\bm S_z|\bm S_x) = 0.5$.
For large $\sigma_2$ 
${\rm Var}(\bm S_z|\bm S_x)$
towards to 1/4 as a result of 
the weak interaction in the 
second measurement.
Likewise, for $x_2 > 0$,
${\rm Var}(\bm S_z|\bm S_x)$ 
ranges from 1/8 to 1/4 
when $\sigma_2$ goes from
strong to weak interaction, 
respectively. 
(b) Plot of ${\rm Var}(\bm S_z|\bm S_x)$
as a function of $x_2$ for 
$\sigma_2\in\{0.1, 0.5, 1.0, 2.0\}$.
(c) Plot of ${\rm Var}(\bm S_z|\bm S_x)$
as a function of $\sigma_2$ for 
$x_2\in\{-1.0, -0.5, 0.0, 0.5, 1.0\}$.
}
\label{fig4}
\end{figure*} 

Next, we examine the conditional variance 
of the first measurement
$\bm S_z$ conditioned on 
the presence of the 
second measurement
$\bm S_x$. It reads
\begin{align}\label{dAB}
{\rm Var}(\bm S_z|\bm S_x) = 
\dfrac{1}{4} \dfrac{(s_1-1)s_2}
{s_1s_2-2},
\end{align}
where $s_1=1+\exp({\frac{1}{8\sigma_1^2}})$
and $s_2=1+\exp({\frac{x_2}{\sigma_2^2}})$.
There is an interplay 
between $\sigma_1$ and $\sigma_2$.
When $\sigma_1\to 0$ 
(the first measurement is strong), then
$s_1\to \infty$,
it will dominate the result,
and thus the variance 
${\rm Var}(\bm S_z|\bm S_x) 
= 1/4 = {\rm Var}(\bm S_z)_{\rho_0}$,
regardless $s_2$. 
In this case, the second measurement 
does not disturb the first
because ${\rm Var}(\bm S_z|\bm S_x)
={\rm Var}(\bm S_z)_{\rho_0}$.
Inversely, when $\sigma_1\to \infty$ 
(the first measurement is weak), then
$s_1\approx 2$, and thus 
${\rm Var}(\bm S_z|\bm S_x) \approx
\frac{1}{8}\frac{s_2}{s_2-1}
=\frac{1}{8}[1+{\rm exp}({-\frac{x_2}{\sigma_2^2}})].
$
In this case, the first measurement 
is disturbed 
by the second via $x_2$
and $\sigma_2$. 
The results are shown in Fig.~\ref{fig4}.
Unlike the previous case, 
the variance ${\rm Var}(\bm S_z|\bm S_x)$ 
is not symmetrical via the outcome $x_2$. 
When $x_2 = 0$, 
${\rm Var}(\bm S_z|\bm S_x)$ reaches 1/4, 
indicating that the first measurement 
is not affected by the second.
When $x_2 < 0$, it will be 
strongly disturbed, wherein 
$1/4 < {\rm Var}(\bm S_z|\bm S_x) < \infty$.
Here, ${\rm Var}(\bm S_z|\bm S_x) \to \infty$
when $\sigma_2\to 0$
(we show the result up to 
${\rm Var}(\bm S_z|\bm S_x) = 0.5$).
When $x_2 > 0$, the variance 
is bound in the range
$1/8 \le {\rm Var}(\bm S_z|\bm S_x) < 1/4$
when $\sigma_2$ goes from 
strong to weak interaction.

Finally, when both measurements are weak,
such as sequential weak measurements
\cite{PhysRevA.76.062105,
PhysRevLett.117.170402,
PhysRevA.96.052123,
Brodutch2017,
PhysRevA.98.012117,
Kim2018,Chen:19,Pfender2019},
the QBA effects are eliminated,
i.e., ${\rm Var}(\bm S_x|\bm S_z)\to 
{\rm Var}(\bm S_x)_{\rho_0}$,
and ${\rm Var}(\bm S_z|\bm S_x)
\to {\rm Var}(\bm S_z)_{\rho_0}$.

\subsection{Conditonal stronger uncertainty relations} 
We study the MPUR in the spin model
with conditional measurement. 
The initial system state is 
$|\psi\ra = |+\ra$,
an eigenstate of $\bm S_x$.
We thus choose 
$|\psi^\perp\ra = 
(\bm S_z - \la\bm S_z\ra)|
+\ra/\Delta(\bm S_z)
=|-\ra$
\cite{PhysRevLett.113.260401},
and obtain
$|\psi^\perp_{\bm S_z+\bm S_x}\ra 
= |-\ra$.
Then, $\mc{R}_{\rm a}$ and  
$\mc{R}_{\rm b}$ 
in Eq.~\eqref{eq:MPAB} yield
\begin{align}
        \mc{R}_{\rm a} = 
       \dfrac{1}{4}, \text {and }
        \mc{R}_{\rm b} =
       \dfrac{1}{8}.
\end{align}
The MPUR
for the system before 
the measurements reads
\begin{align}\label{eq:MPUR1}
{\rm Var}(\bm S_z)_{\rho_0} + 
{\rm Var}(\bm S_x)_{\rho_0} = \dfrac{1}{4}
= \max(\mc{R}_{\rm a},
\mc{R}_{\rm b}),
\end{align}
in which reaches the lower bound
in Eq.~\eqref{eq:MUP}.

Next, we consider the MPUR for the
conditional variances. Since 
${\rm Var}(\bm S_z|\bm S_x)$
and ${\rm Var}(\bm S_x|\bm S_z)$
can be archived individually, we
obtain the inequality 
${\rm Var}(\bm S_z|\bm S_x)
+{\rm Var}(\bm S_x|\bm S_z) 
\ge \frac{1}{8}$.
The equality is attainable when 
${\rm Var}(\bm S_z|\bm S_x) = \frac{1}{8}$
and ${\rm Var}(\bm S_x|\bm S_z) = 0$,
as previously discussed. 
This is the lower bound for MPUR
in the
conditional 
sequential measurements, 
resulting in enhanced precision 
measurements and reduced QBA effects.

\section{$N$-sequential measurements}
\label{seciv}
We extend the framework to 
$N$-sequential measurements.
Let $\bm U_k = e^{-i\bm A_k\otimes \bm p_k}$
be the evolution operator of the
$k^{\rm th}$ measurement,
where $\bm A_k = 
\int a_k|a_k\ra\la a_k| 
\ {\rm d}a_k$ is the 
system operator,
and $\bm p_k$ is the $k^{\rm th}$ 
probe operator.
Let $\bm\Omg_{k}$ be the 
measurement operator
of the $k^{\rm th}$ measurement,
we have
\begin{align}\label{eq:npovm}
\bm \Omg_{k} = \int\psi_{k}(x_k-a_k)
|a_k\ra\la a_k|\ {\rm d}a_k.
\end{align}
The system state after $k$ 
measurements is given by
\begin{align}\label{eq:rhoN}
\bar\rho_k = \bm\Omg_{k}\bar\rho_{k-1}
\bm\Omg^\dagger_{k},
\ \text { for } \ k = 1, \cdots, N.
\end{align}
The probability of obtaining 
outcome $x_k$ 
conditional on
the rest of other 
outcomes is given by
\begin{align}\label{eq:conN}
p(x_k|\bar{x}_k)
&= \dfrac{{\rm Tr}[\bar\rho_k\bm E]}
{\int {\rm Tr}[\bar\rho_k\bm E]\ {\rm d}x'_k},
\end{align}
where $\bar{x}_k = \{x_1,\cdots,x_N\}
\backslash x_k$ stands for the rest of 
other outcomes, i.e.,
a set of all elements
from $x_1$ to $x_N$
except $x_k$, and
\begin{align}\label{eq:EN}
\bm E = \bm\Omg^\dagger_{k+1}
\bm\Omg^\dagger_{k+2}\cdots
\bm\Omg^\dagger_{N}
\bm\Omg_{N}
\bm\Omg_{N-1}\cdots
\bm\Omg_{k+1}.
\end{align}
See the proof in App.~\ref{appE}.
This result is similar
to the one given in Ref.~\cite{Bao2020}.
However, in Ref.~\cite{Bao2020},
the authors approximate 
the Gaussian distribution 
of the conditional probability to deduce the 
variance and 
use a hypothesis projective measurement
at the $k^{\rm th}$ measurement.
Here, we derive the exact variance
from the expectation value of the 
position operator $\bm x_k$.
At first, the conditional expectation value 
is given by
\begin{align}\label{eq:expx2}
E(\bm {x}_{k}|
{\bar{x}_k}) = 
\int x_k\ p
(x_k|\bar{x}_k)
 {\rm d}x_k.
\end{align}
Then, the conditional variance is
expressed as
\begin{align}\label{condx2N}
\notag {\rm Var}(\bm {x}_k|\bar{x}_k)
&= 
E(\bm x_k^2|{\bar{x}_k}) - 
[E(\bm {x}_{k}|{\bar{x}_k})]^2\\
&= \sigma_k^2 + {\rm Var} (\bm A_k\big|
\bar{\bm A}_k) ,
\end{align}
where $\bar{\bm A}_k
= \{\bm A_1,\cdots,\bm A_N\}\backslash \bm A_k$.
We thus can extract the conditional 
variance 
${\rm Var}(\bm A_k|\bar{\bm A}_k)$.

For example, in App.~\ref{appE}, we
provide a useful Mathematica code
for calculating the conditional variance
${\rm Var}(\bm x_2|x_1, x_3, x_4)$
of 4-sequential measurements:
$\bm S_z \to \bm S_x \to \bm S_x \to \bm S_z$.

\section{Conclusion and outlook}\label{secv}
We studied the impact of quantum backaction
(QBA) on sequential measurements of
incompatible operators.
We considered the indirect 
measurement of observables $\bm A$
and $\bm B$ using two quantum probes
$\mathcal{P}_1$ and $\mathcal{P}_2$
in the joint and conditional measurement models.
In the joint measurement, 
only the QBA caused by the 
first measurement is presented 
due to the influence of former 
measurements on later ones.
Otherwise, in the conditional 
measurement approach,
there is a mutual effect
in sequential measurements
expressed through the QBAs of
the first and second measurements.
We illustrated the results in a spin system,
showing the dependence of 
variance in subsequent measurements 
on prior measurements and vice versa.
This observation results from the interplay between
the measurement strengths of 
the first and second measurements.
We further derived a new bound 
for the stronger uncertainty relation 
with conditional measurements.
The framework 
was then extended to $N$-sequential measurements.

Finally, we remark that these findings 
can be demonstrated 
using current technologies, 
such as trapped ions, macroscopic vapor cells,
and nuclear magnetic resonance. 
As we showed theoretically, 
the measurement operators $\bm M_m$
and $\bm N_n$ can be implemented in
the phase-space representation of the driven field
(e.g., light, magnetic fields, etc),
and the measurement results can be inferred from
the position-shift of the final state.
Potential future research could include 
controlling and eliminating the QBA effects,
experimental verifying the stronger uncertainty relation.


\newpage

\begin{widetext}
\appendix
\setcounter{equation}{0}
\renewcommand{\theequation}{A.\arabic{equation}}
\section{Final joint state in the 
joint measurement model}\label{appA}

In this section, we derive the final state
of sequential measurements 
in the joint measurement model:
\begin{align}\label{app:fn}
\rho_{\rm fin} = \bm U_2\bm U_1\rho_{\rm int}
\bm U_1^\dagger \bm U_2^\dagger.
\end{align}
Inserting $\bm I_1 = \int |m\ra\la m| {\rm d}m$
and $\bm I_2 = \int |n\ra\la n| {\rm d}n$, we have
\begin{align}\label{app:fn1}
\notag
\rho_{\rm fin} = 
\int |n\ra\la n|
&\bm U_2
\int |m\ra\la m|
\bm U_1
\Bigl(\rho_0 
\otimes |\psi_1(x)\ra\la\psi_1(x)|
\otimes |\psi_2(x)\ra\la\psi_2(x)|\Bigr)
\bm U_1^\dagger \\
&\times \int |m'\ra\la m'|
\bm U_2^\dagger
\int |n'\ra\la n'|
{\rm d}m{\rm d}m'{\rm d}n{\rm d}n'.
\end{align}
%
Set $\bm M_m = 
\bigl(\bm I_{\mc{S}}\otimes \la m|\bigr)
\bm U_1
\bigl(\bm I_{\mc{S}} \otimes |\psi_1(x)\ra
\bigr)$
is the measurement operator
for the first measurement, 
where $\bm I_{\mc{S}}$ is an identity operator
in the system space and can be expressed in terms of
eigenstates of $\bm A$ (or $\bm B$), i.e., 
$\bm I_{\mc{S}} = \int |a\ra\la a| {\rm d}a$.
Particularly, we have
\begin{align}\label{app:Mm}
\notag \bm M_m &= 
\bigl(\bm I_{\mc{S}}\otimes\la m|\bigr)
e^{-i\bm A\otimes \bm p_1}
\int |a\ra\la a|\otimes 
|\psi_1(x)\ra
{\rm d}a\\
\notag &= 
\bigl(\bm I_{\mc{S}}\otimes\la m|\bigr)
\int |a\ra\la a|\otimes 
e^{-ia\bm p_1}|\psi_1(x)\ra
{\rm d}a\\
&= 
\int |a\ra\la a|\otimes
\la m|e^{-ia\bm p_1}|\psi_1(x)\ra
{\rm d}a.
\end{align}
Submitting $|\psi_1(x)\ra = \int \psi_1(x)|x\ra
{\rm d}x$ into Eq.~\eqref{app:Mm}, we obtain
\begin{align}\label{app:Mmfin}
\notag \bm M_m &= 
\int \Bigl[
|a\ra\la a|\otimes
\la m|e^{-ia\bm p_1}\int \psi_1(x)|x\ra
{\rm d}x\Bigr]
{\rm d}a\\
&= 
\int \psi_1(m-a)
|a\ra\la a| {\rm d}a.
\end{align}
The measurement operator $\bm M_m$
satisfies the completeness relation as
$\int \bm M_{m}^\dagger \bm M_{m} {\rm d}m
= \bm I$:
\begin{align}\label{eq:povm}
\notag \int \bm M_{m}^\dagger \bm M_{m} {\rm d}m 
	&= \int 
	\Big[
	 \int \psi_1^*(m-a)|a\ra\la a|{\rm d}a \times 
\int \psi_1(m-a')|a'\ra\la a'|{\rm d}a'
	\Big] {\rm d}m \\
\notag& = \int \Big[\int
	\psi_1^*(m-a)\psi_1(m-a) {\rm d}m\Big]
	|a\ra\la a| {\rm d}a\\
	&=\bm I.
\end{align}

Similarly,  for the second measurement, we define 
a measurement operator
$\bm N_n = \bigl(\bm I_{\mc{S}}\otimes \la n|\bigr)
\bm U_2|
\bigl(\bm I_{\mc{S}} \otimes \psi_2(x)\ra
\bigr)$,
which particularly gives
\begin{align}\label{app:Mmfin}
 \bm N_n = 
\int \psi_2(n-b)
|b\ra\la b| {\rm d}b,
\end{align}
and obeys the relation
$\int \bm N_{n}^\dagger \bm N_{n} {\rm d}n
= \bm I.$
Inserting $\bm M_m$ and $\bm N_n$ into
Eq.~\eqref{app:fn1}, we obtain
\begin{align}\label{app:fn2}
\rho_{\rm fin} &= 
\iiiint \bm N_n\bm M_m
\rho_0 
\bm M^\dagger_{m'}
\bm N^\dagger_{n'}\
 |m\ra\la m'|
\otimes
|n\ra\la n'|
{\rm d}m{\rm d}m'{\rm d}n{\rm d}n',
\end{align}
which is given in 
Eq.~\eqref{fin_state} 
in the main text.

\setcounter{equation}{0}
\renewcommand{\theequation}{B.\arabic{equation}}
\section{Final joint probes state 
in the joint measurement model}\label{appB}
Here, we derive the joint state of
$\mc{P}_1\otimes\mc{P}_2$
by partially tracing out the system space
\begin{align}\label{app:pointersstate}
\eta = {\rm Tr}_{\mc{S}} 
[\rho_{\rm fin}].
\end{align}
We first consider the cyclic property 
of the trace as 
${\rm Tr}_{\mc{S}}
\bigl[\bm N_n\bm M_m\rho_0
\bm M^\dagger_{m'}
\bm N^\dagger_{n'}\bigr]
=
{\rm Tr}_{\mc{S}}
\bigl[\bm M_m\rho_0
\bm M^\dagger_{m'}
\bm N^\dagger_{n'}\bm N_n
\bigr],
$
where
\begin{align}\label{app:outter}
\notag  \bm N^\dagger_{n'} \bm N_{n}
& = 
\int \psi_2^*(n'-b')|b'\ra\la b'| 
\int \psi_2(n-b)|b\ra\la b| {\rm d}b {\rm d}b'
\\
&=
\int \psi_2(n-b)\psi_2^*(n'-b)
|b\ra\la b|
{\rm d}b,
\end{align}
and
\begin{align}\label{app:inter}
\notag \bm M_m\rho_0 \bm M^\dagger_{m'}
& = 
\Bigl[\int \psi_1(m-a)|a\ra
\la a| {\rm d}a\Bigr] \times
\rho_0 
\times\Bigl[\int \psi_1^*(m'-a')|a'\ra
\la a'| {\rm d}a'\Bigr]\\
& = 
\iint |a\ra\la a|
\rho_0 |a'\ra\la a'| \times
\psi_1(m-a)\psi_1^*(m'-a')
{\rm d}a {\rm d}a'.
\end{align}
%
From Eqs.~(\ref{app:outter}, 
\ref{app:inter}), we have
\begin{align} 
{\rm Tr}_{\mc{S}}
\bigl[\bm M_m\rho_0
\bm M^\dagger_{m'}
\bm N^\dagger_{n'}\bm N_n\bigr] = 
\iiint \la b|a\ra
\la a|\rho_0| a'\ra\la a'|b\ra \times 
 \psi_1(m-a) \psi_1^*(m'-a')\ \times 
\psi_2(n-b)\psi_2^*(n'-b)
{\rm d}a {\rm d}a' {\rm d}b.
\end{align}
Submitting it into Eq.~\eqref{app:fn2}
and taking trace over
system $\mc{S}$,
we get  Eq.~\eqref{app:pointersstate},
which explicitly gives
\begin{align}\label{app:pse}
\eta = 
		\iiint & \la b|a\ra
		\la a|\rho_0| a'\ra
		\la a'|b\ra\times
		|\psi_1(m-a)\ra
		\la\psi_1(m'-a')|  \otimes 
		|\psi_2(n-b)\ra
		\la\psi_2(n'-b)|
		{\rm d}a {\rm d}a' {\rm d}b,
\end{align}
where we used
$\int \psi_1(m-a)|m\ra {\rm d}m =
|\psi(m-a)\ra$ 
and similarly for these others.
Noting that this formula
does not explicitly depend on $m$, 
we thus can replace $m$ by $x$, 
and recast Eq.~\eqref{app:pse} by
\begin{align}\label{app:pser}
\eta = 
\iiint \la b|a\ra
\la a|\rho_0| a'\ra
\la a'|b\ra\times
|\psi_1(x-a)\ra
\la\psi_1(x-a')| 
\otimes |\psi_2(x-b)\ra
\la\psi_2(x-b)|
{\rm d}a {\rm d}a' {\rm d}b,
\end{align}
which is shown in Eq.~\eqref{joint_state2} 
in the main text.

\setcounter{equation}{0}
\renewcommand{\theequation}{C.\arabic{equation}}
\section{Variances in the joint measurement model}\label{appC}
\subsection{Derivation of the variance ${\rm Var}(\bm {x}_1)_\eta$}
\label{appC1}
We calculate 
the expectation value of 
$\bm x_1$ 
in Eq. (7) in the main text 
as
\begin{align}\label{app:fe}
\notag \la\bm x_1\otimes \bm I_2\ra
& = {\rm Tr}_{\mc{P}_1\mc{P}_2}
\bigl[(\bm x_1\otimes \bm I_2)\eta\bigr]\\
& =\iiint \la b|a\ra
\la a|\rho_0| a'\ra
\la a'|b\ra \times
\la\psi_1(x-a')|\bm x_1|\psi_1(x-a)\ra
{\rm d}a {\rm d}a' {\rm d}b.
\end{align}
Wherein, we have
\begin{align}\label{trb}
\notag\int  \la b|a\ra
\la a|\rho_0| a'\ra
\la a'|b\ra
{\rm d}b 
&= {\rm Tr}_{\mc{S}}
\Bigl[|a\ra
\la a|\rho_0| a'\ra
\la a'|\Bigr]\\
& = {\rm Tr}_{\mc{S}}
\bigl[|a\ra
\la a|\rho_0\bigr],
\end{align}
where we used the cyclic property 
of the trace, then, only $a = a'$ is non-vanish,
and
\begin{align}\label{app:num_1}
\notag \la\psi_1(x-a')|\bm x_1|\psi_1(x-a)\ra 
&=\Bigl[\int\Bigl(\dfrac{1}{2\pi\sigma_1^2}\Bigr)^{1/4}
e^{-\frac{(x'-a')^2}{4\sigma_1^2}}\la x'| {\rm d}x'\Bigr] 
\times \Bigl[\int x_1 |x_1\ra\la x_1| {\rm d}x_1\Bigr]
\times \Bigl[\int\Bigl(\dfrac{1}{2\pi\sigma_1^2}\Bigr)^{1/4}
e^{-\frac{(x-a)^2}{4\sigma_1^2}} |x\ra {\rm d}x\Bigr]\\
\notag&=\dfrac{a+a'}{2}e^{-\frac{(a-a')^2}{8\sigma_1^2}}\\
&= a \ \text{ for all } a = a'.
\end{align}
Then, the expectation value \eqref{app:fe} is recast as
\begin{align}\label{app:numre}
\notag\la\bm x_1\otimes \bm I_2\ra
&=\int a \ {\rm Tr}_{\mc{S}}
\bigl[|a\ra
\la a|\rho_0\bigr]
{\rm d}a\\
&= {\rm Tr}_{\mc{S}}
[\bm A\rho_0].
\end{align}
Obviously, the expectation value
obtained from the first measurement 
does not depend on the second measurement. 

We next calculate 
$\la(\bm x_1\otimes \bm I_2)^2\ra$ for
the non-vanish terms, i.e., $a = a'$. 
We first evaluate 
\begin{align}\label{app:num_xp2}
\notag \la\psi_1(x-a)|\bm x_1^2|\psi_1(x-a)\ra 
&=\Bigl[\int\Bigl(\dfrac{1}{2\pi\sigma_1^2}\Bigr)^{1/2}
e^{-\frac{(x'-a)^2}{4\sigma_1^2}}\la x'| {\rm d}x'\Bigr] \times
\Bigl[\int x_1^2 |x_1\ra\la x_1| {\rm d}x_1\Bigr]
\times
\Bigl[\int\Bigl(\dfrac{1}{2\pi\sigma_1^2}\Bigr)^{1/2}
e^{-\frac{(x-a)^2}{4\sigma_1^2}} |x\ra {\rm d}x\Bigr] \\
&= \sigma_1^2 + a^2.
\end{align}
Combining with Eq.~\eqref{trb},
we obtain
\begin{align}\label{app:x12}
\notag \la(\bm x_1\otimes \bm I_2)^2\ra
&= \int (\sigma_1^2 + a^2) \ {\rm Tr}_{\mc{S}}
\bigl[|a\ra
\la a|\rho_0\bigr]
{\rm d}a\\
&=\sigma_1^2 + 
{\rm Tr}_{\mc{S}}
\bigl[\bm A^2\rho_0\bigr].
\end{align}
Then, the variance is
\begin{align}\label{app:var1}
\notag{\rm Var}(\bm x_1)_{\eta} & 
= \la(\bm x_1\otimes \bm I_2)^2\ra
- \la\bm x_1\otimes \bm I_2\ra^2\\
& = \sigma_1^2+{\rm Var}(\bm A)_{\rho_0},
\end{align}
where ${\rm Var}(\bm A)_{\rho_0}
= {\rm Tr}_{\mc{S}}[\bm A^2\rho_0] - 
({\rm Tr}_{\mc{S}}[\bm A\rho_0])^2$.
This result is casually trivial,
where the second measurement does not affect
the first one.

\subsection{Derivation of the variance ${\rm Var}(\bm x_2)_\eta$}
\label{appC2}
We next examine the backaction effect 
of the first measurement affects the
second measurement. 
We start from the expectation value:
\begin{align}\label{app:nums2}
\notag \la\bm I_1\otimes \bm x_2\ra 
& =  {\rm Tr}_{\mc{P}_1\mc{P}_2}
\bigl[(\bm I_1\otimes \bm x_2)\eta\bigr] \\
& =\iiint \la b|a\ra
\la a|\rho_0| a'\ra
\la a'|b\ra \times 
\la\psi_1(x-a')|\psi_1(x-a)\ra\ \times 
\la\psi_2(x-b)|\bm x_2|\psi_2(x-b)\ra
{\rm d}a {\rm d}a' {\rm d}b.
\end{align}
The middle term is
\begin{align}\label{app:mid}
\la\psi_1(x-a')|\psi_1(x-a)\ra =
e^{-\frac{(a-a')^2}{8\sigma_1^2}}.
\end{align}
The last term is (similar to \eqref{app:num_1})
\begin{align}\label{app:mid}
\la\psi_2(x-b)|\bm x_2|\psi_2(x-b)\ra =
b.
\end{align}
Then, Eq.~\eqref{app:nums2} explicitly gives
\begin{align}\label{app:nums2f}
\notag \la\bm I_1\otimes \bm x_2\ra
&=\iiint b\la b|a\ra
\la a|\rho_0| a'\ra
\la a'|b\ra
  e^{-\frac{(a-a')^2}{8\sigma_1^2}}
{\rm d}a {\rm d}a' {\rm d}b\\
& = {\rm Tr}_{\mc{S}}\Bigl[
\bm B\rho_1\Bigr],
\end{align}
where
\begin{align}\label{app:rho1}
\notag \rho_1 &= {\rm Tr}_{\mc{P}_1}
\Bigl[
e^{-i{\bm A}\otimes\bm p_1}
    \Bigl(\rho_0\otimes 
            |\psi_1(x)\ra\la\psi_1(x)|
    \Bigr)
e^{i{\bm A}\otimes\bm p_1}           
\Bigr]\\
&= \iint |a\ra\la a|
\rho_0
|a'\ra\la a'|
e^{-\frac{(a-a')^2}{8\sigma_1^2}}
{\rm d}a {\rm d}a',
\end{align}
is the system state after the first interaction.
For weak measurement, 
$\sigma_1 \gg |a-a'|$, 
i.e., $e^{-\frac{(a-a')^2}
{8\sigma_1^2}} \approx 1$,
we have
$\la\bm I_1\otimes \bm x_2\ra
\approx {\rm Tr}_{\mc{S}}[\bm B\rho_0]$.
%
We next calculate the term 
$\la(\bm I_1\otimes \bm x_2)^2\ra$ 
and obtain
\begin{align}\label{app:avw2f}
 \la(\bm I_1\otimes \bm x_2)^2\ra
=\sigma_2^2 + 
{\rm Tr}_{\mc{S}}[\bm B^2\rho_1]
\end{align}
Finally, the variance is given by
\begin{align}\label{app:varx2}
{\rm Var}({\bm x_2})_{\eta} 
= \sigma_2^2 + {\rm Var}(\bm B)_{\rho_1},
\end{align}
where $ {\rm Var}(\bm B)_{\rho_1} = 
 {\rm Tr}_{\mc{S}}[\bm B^2\rho_1] 
- ( {\rm Tr}_{\mc{S}}[\bm B\rho_1] )^2
$ and particularly gives
\begin{align}\label{app:DelBp}
{\rm Var}(\bm B)_{\rho_1} = 
\int b^2\la b|\rho_1|b\ra
{\rm d}b
- \Bigl[
\int b\la b|\rho_1|b\ra
{\rm d}b
\Bigr]^2.
\end{align}

For the weak measurement, 
${\rm Var}(\bm B)_{\rho_1} = {\rm Var}(\bm B)_{\rho_0}$,
which implies that the weak measurement 
will not affect the later measurement results.

\setcounter{equation}{0}
\renewcommand{\theequation}{D.\arabic{equation}}
\section{Conditional measurement model}\label{appD}

We calculate the conditional probability
\begin{align}\label{app:cp}
p(x_2\big|x_1) = 
\dfrac{
{\rm Tr}\Bigl[\bar{\rho}_1(x_1)
\bm E(x_2)\Bigr]}
{\bigintssss
{\rm Tr}\Bigl[\bar{\rho}_1(x_1)
\bm E(x'_2)\Bigr] {\rm d}x'_2}.
\end{align}
Using $\bm E(x_2) = \bm N_{x_2}^\dagger \bm N_{x_2}$,
we have
\begin{align}\label{app:NrhopN}
\notag 
{\rm Tr}\Bigl[\bar{\rho}_1(x_1)
\bm E(x_2)\Bigr] &= 
{\rm Tr}\Bigl[ \bm N_{x_2}\bar{\rho}_1(x_1)
\bm N^\dagger_{x_2}\Bigr]\\
\notag &={\rm Tr}\Bigl[ \int
\psi_2(x_2-b)|b\ra\la b|{\rm d}b\times
\bar\rho_1(x_1)
\times \int
\psi_2^*(x_2-b')|b'\ra\la b'|{\rm d}b' \Bigr]\\
&=\int \big|\psi_2(x_2-b)\big|^2
\la b|\bar\rho_1(x_1)|b\ra {\rm d}b.
\end{align}
Submitting to Eq.~\eqref{app:cp},
we obtain 
\begin{align}\label{app:cp1}
p(x_2\big|x_1) 
=\dfrac{
\bigintssss \big|\psi_2(x_2-b)\big|^2\la b|\bar{\rho}_1
| b\ra\ {\rm d}b}
{\bigintssss\Bigl[\bigintssss
\big|\psi_2(x'_2-b)\big|^2\la b|\bar{\rho}_1
|b\ra\ {\rm d}b\Bigr] {\rm d}x'_2},
\end{align}
as shown in Eq. (14) 
in the main text.

\setcounter{equation}{0}
\renewcommand{\theequation}{E.\arabic{equation}}
\section{$N$-sequential measurements}\label{appE}
Here, we prove Eq. (35) 
in the main text. Under $N$-sequential measurements,
the joint probability is given by
\begin{align}\label{app:jp}
\notag p(x_1,\cdots,x_N)
&={\rm Tr}_{\mc{S}}\bigl[
\bm\Omg_N\cdots\bm\Omg_{k+1}
\bar\rho_{k}\bm\Omg_{k+1}^\dagger\cdots
\bm\Omg_N^\dagger\bigr]\\
\notag&= {\rm Tr}_{\mc{S}}\bigl[
\bar\rho_{k}
\bm\Omg_{k+1}^\dagger\cdots
\bm\Omg_N^\dagger\bm\Omg_N
\cdots\bm\Omg_{k+1}\bigr]\\
&= {\rm Tr}_{\mc{S}}\bigl[
\bar\rho_{k} E\bigr],
\end{align}
where we used $\bar\rho_{k} = \bm\Omg_k
\bar\rho_{k-1}\bm\Omg_k^\dagger$,
and $E = \bm\Omg_{k+1}^\dagger\cdots
\bm\Omg_N^\dagger\bm\Omg_N
\cdots\bm\Omg_{k+1}$.
Then, the conditional probability is defined as
\begin{align}\label{app:conN0}
\notag p(x_k\big|
\bar{x}_k
) &\equiv
\dfrac{ p(x_1,\cdots,x_N)}
{\int p(x_1,\cdots,x_N) {\rm d}x'_k}\\
&=\dfrac{{\rm Tr}_{\mc{S}}\bigl[
\bar\rho_{k} \bm E\bigr]}
{\int {\rm Tr}_{\mc{S}}\bigl[
\bar\rho_{k} \bm E\bigr] {\rm d}x'_k},
\end{align}
where $\bar{x}_k = \{x_1,\cdots,x_N\}
\backslash x_k$.

Next, substituting $\bm\Omg_k
= \int\psi_k(x_k-a_k)|a_k\ra\la a_k|\
{\rm d}a_k$ into $\bar\rho_k$, we obtain
\begin{align}\label{app:rhok}
\notag \bar\rho_k &= 
\Bigl[
    \int\psi_k(x_k-a_k)|a_k\ra\la a_k|
    {\rm d}a_k
\Bigr] \times
    \bar\rho_{k-1}\ \times 
\Bigl[
    \int\psi_k^*(x_k-a'_k)|a'_k\ra\la a'_k|
    {\rm d}a'_k
\Bigr],\\
& =  
 \iint\Bigl[\psi_k(x_k-a_k)\psi_k^*(x_k-a'_k)
 |a_k\ra\la a_k| \bar\rho_{k-1}
 |a'_k\ra\la a'_k|\Bigr]
{\rm d}a_k
 {\rm d}a'_k.
\end{align}
Submitting Eq.~\eqref{app:rhok} into 
Eq.~\eqref{app:conN0}, we have
\begin{align}\label{app:conN}
p(x_k\big|\bar{x}_k)
 = \dfrac{\iint \psi_k(x_k-a_k)\psi_k^*(x_k-a_k')
\la a_k|\bar\rho_{k-1}|a_k\ra
\la a_k'|\bm E|a_k\ra\ {\rm d}a_k {\rm d}a'_k}
{\int \bigl[\iint \psi_k(x_k-a_k)\psi_k^*(x_k-a_k')
\la a_k|\bar\rho_{k-1}|a_k\ra
\la a_k'|\bm E|a_k\ra\ {\rm d}a_k {\rm d}a'_k\bigr] {\rm d}x_k'}.
\end{align}

A useful Mathematica code for calculating 
the conditional variance
is given is Listing 1.

\begin{lstlisting}[
label = {list:1}
language=Mathematica,
caption={
Mathematica code
for calculating the conditional variance
${\rm Var}(\bm x_2|x_1, x_3, x_4)$
of 4-sequential measurements:
$\bm S_z \to \bm S_x \to \bm S_x \to \bm S_z$.
},
mathescape=true,
aboveskip=\medskipamount]
$\textcolor{cyan}{\text{(* Eigenstates and eigenvectors *)}}$
ez1 = 0.5;
ex2 = -0.5;
MatrixForm[vecz1 = {{1},{0}}];
MatrixForm[vecz2 = {{0},{1}}];

ex1 = 0.5;
ex2 = -0.5;
MatrixForm[vecx1 = 1/Sqrt[2]*{{1},{1}}];
MatrixForm[vecx2 = 1/Sqrt[2]*{{1},{-1}}];

$\textcolor{cyan}{\text{(* Measurement operators *)}}$
$\psi$[$\sm$,$\xm$,$\am$]:=$\Bigl(\dfrac{1}{2\pi*\sn^2}\Bigr)^{1/4}$*Exp[-$\dfrac{(\xn-\an)^2}{4\sn^2}$];$\vspace{0.5cm}$
$\Omg$[$\sm$,$\xm$,$\emm$,$\vm$]:= $\vn^\dagger.\vn*\psi$[$\sn$,$\xn$,$\an$];
$\Omg$Z[$\sm$,$\xm$]:= $\Omg$[$\sn$,$\xn$,ez1,vecz1]+$\Omg$[$\sn$,$\xn$,ez2,vecz2];
$\Omg$X[$\sm$,$\xm$]:= $\Omg$[$\sn$,$\xn$,ex1,vecx1]+$\Omg$[$\sn$,$\xn$,ex2,vecx2];

$\textcolor{cyan}{\text{(* Evolution system state *)}}$
MatrixForm[$\rho0$ = {{0.5, 0.5},{0.5, 0.5}}];
$\rho1$:= FullSimplify[$\Omg$Z[$\bsOn$,$\bxOn$].$\rho0$.$\Omg$Z[$\bsOn$,$\bxOn$]];
$\rho2$:= FullSimplify[$\Omg$X[$\bsTn$,$\bxTn$].$\rho1$.$\Omg$X[$\bsTn$,$\bxTn$]];
fE:=$\Omg$X[$\bsTrn$,$\bxTrn$].$\Omg$Z[$\bsFn$,$\bxFn$].$\Omg$Z[$\bsFn$,$\bxFn$].$\Omg$X[$\bsTrn$,$\bxTrn$];

$\textcolor{cyan}{\text{(* Conditional probability *)}}$
Num = Tr[$\rho2$.fE];
Den= Integrate[Num, {$\xTn$,-$\infty,\infty$}];
Pr = FullSimplify[Num / Den];

$\textcolor{cyan}{\text{(* Conditional variance *)}}$
Average = Simplify[Integrate[$\xTn$*Pr, {$\xTn$,-$\infty,\infty$}]];
Average2 = Simplify[Integrate[$\xTn^2$*Pr, {$\xTn$,-$\infty,\infty$}]];
Variances = FullSimplify[Average2 - Average$^2$];
\end{lstlisting}

\end{widetext}

\bibliography{refs}
\end{document}